\documentclass[prl,aps,amsfonts,amssymb,amsmath,showpacs,twocolumn]{revtex4}

\usepackage{times}
\usepackage{graphicx}
\usepackage{bm}

\begin{document}

\title{Wave turbulence theory for gravitational waves in general relativity: The Space-Time Kolmogorov spectrum. }

\author{Sergio Rica\thanks{sergio.rica@uai.cl}}
\affiliation{Facultad de Ingenier\'ia y Ciencias, Universidad Adolfo Ib\'a\~nez, Avda. Diagonal las Torres 2640, Pe\~nalol\'en, Santiago, Chile}

\begin{abstract}
The recent observation of gravitational waves, stimulates the question of the longtime evolution of the space-time fluctuations. Gravitational waves interact themselves through the nonlinear character of Einstein's equations of general relativity. This nonlinear wave interaction allows the spectral energy transfer from mode to mode. According to the wave turbulence theory, the weakly nonlinear interaction of gravitational waves leads to the existence of an irreversible kinetic regime that dominates the longtime evolution. The resulting kinetic equation suggests the existence of an equilibrium wave spectrum and the existence of a  non-equilibrium Kolmogorov-Zakharov spectrum for spatio-temporal fluctuations.  Evidence of these solutions extracted in the fluctuating signal of the recent observations will be discussed in the paper. Probably, the present results would be pertinent in the new age of development of gravitational astronomy, as well as, in new tests of General Relativity.
\end{abstract}


\date{\today}

\maketitle

{\it i) Introduction-} Recent evidence of the existence of gravitational waves \cite{abbott,abbott.new}, predicted by Einstein's theory of gravitation, motivates the question of the longtime evolution of the space-time fluctuations. The physics we have in mind concerns the temporal evolution of the space-time metric of an ensemble of suitable stochastic initial conditions in vacuum. As a consequence of Einstein's equations of general relativity, spatio-temporal fluctuations propagate and interact themselves as time goes.  In this way the spatiotemporal ripples travel and interact themselves allowing the spectral redistribution of the gravitational wave energy produced by the space-time from one scale to another building-up a coherent spectral energy distribution.

Since the last  50 years, it was established that the long time statistical properties of a random fluctuating wavy system possess a natural asymptotic closure because of the weakly nonlinear wave interactions \cite{hasselmann,benney}.  Indeed this so-called ``wave turbulence theory" has shown to be a powerful method to study the evolution of nonlinear wave systems. 
It results that the longtime dynamics is driven by a kinetic equation for the distribution of spectral densities. This method has been successfully applied for surface gravity waves in fluids \cite{hasselmann,zakgrav66}, surface capillary waves in fluids \cite{zakcap67}, plasma waves \cite{zakplasma67}, sound waves \cite{zs}, nonlinear optics \cite{Dyachenko-92,picozzi}, nonlinear elastic plates \cite{during}, among others. 

The wave turbulence  kinetic equation has similar non-equilibrium properties  than the usual Boltzmann equation for dilute gases, thus it conserves energy, momentum, and exhibits a H-theorem driving the system to equilibrium, characterized by the Rayleigh-Jeans distribution \cite{RayleighJeans}. More important, besides the elementary equilibrium (or thermodynamic) solution, Zakharov has shown \cite{zakplasma67} that power law non-equilibrium solutions also arise, namely the Kolmogorov--Zakharov (KZ) solutions or KZ spectra, which describe the exchange of conserved quantities ({\it e.g.} energy) between large and small length scales.

The present paper considers a stochastic initial value problem for the space-time metric mandated by Einstein's equations of general relativity. The perturbations of the Minkowski metric, $h_{\mu\nu}$, are waves that travel randomly through the universe interacting resonantly among them {\it via} 
the weak nonlinearities inherent to the gravitation theory. The current nonlinear interactions of the Einstein's equations of general relativity are quadratic up to the first non trivial order.  Hence, the wave interaction includes a three wave process. The mathematics beyond the resonant condition is formally identical to the conservation of energy and momentum in a gas of phonons. In this sense gravitational waves are formally equivalent, in the classical limit, to a gas of interacting  phonons with a non-trivial scattering cross-section as in the well-known Landau-Rumer theory \cite{landau}. Indeed, an isolated system evolves from a random initial condition to a situation of statistical equilibrium like a gas of phonon does. We underline that the actual observed regime of gravitational waves is purely classical.

In addition to statistical equilibrium for isolated systems, the wave
 turbulence theory predicts here an energy cascade from a large scale energy source to a 
 small scale typically beyond the applicability of the theory. 
The evidence of a second black hole merger \cite{abbott.new} suggests a peculiar source for gravitational wave energy. The current occurrence of black hole mergers inject gravitational wave energy to space-time mimicking a sea of random waves, this ``stochastic background''  has been studied recently in Ref. \cite{abbott.Stochastic}, and we shall come back to this point at the end.

{\it ii) Basic Equations.-}
The Einstein equations in vacuum reads,
$R_{\mu\nu}=  0.$
As it is very well know these are 10 equations for the metric tensor which also represents 10 unknowns, however, the Einstein's equations are gauge invariant. In the following we shall use the so called harmonic gauge 
$
g^{\mu\nu}\Gamma^\lambda_{ \mu\nu}=0$ \cite{weinberg}.
We shall assume that the metric is a formal series expansion in a small parameter $\epsilon$ which characterizes the amplitude of the wave, (hence, it is small): $g_{\mu\nu}  = \eta_{\mu\nu} + \epsilon h_{\mu\nu} +\dots$,
here $ \eta_{\mu\nu}$ is the Minkowski metric with the signature $(-1,1,1,1)$ \cite{weinberg}. Next, we expand all pertinent quantities, namely, the Christoffel symbol, the Riemann tensor, and, the Ricci tensor in powers of $\epsilon$, as a functions of $h_{\mu\nu}$ and its derivatives.  The resulting Ricci tensor reads $R_{\mu\nu}=\epsilon \, { R^{(1)}}_{\mu\nu}+ \epsilon^2  \,{ R^{(2)}}_{\mu\nu} +\dots$. Similarly, the gauge condition up to first order constrains $h_{\mu\nu}$, by  $ \partial_\mu {h^\mu}_\nu = \frac{1}{2} \partial_\nu h$.  
Under this gauge the first order of the Ricci's tensor reads
${ R^{(1)}}_{\mu\nu}   = 1/2  {\partial^\lambda}_\lambda  h_{\mu\nu}$.
Because, general relativity does not introduce any intrinsic length scale the second order of the Ricci's tensor has only second  order derivative in space-time coordinates and scales as ${ R^{(2)}}_{\mu\nu} \sim \partial_\mu  h^{\lambda\rho}\partial_\nu   h_{\lambda\rho}$  (the full expression of the ${ R^{(2)}}_{\mu\nu}$ may be found in Sec. 7.6 of  \cite{weinberg}).
 
These quantities together with the Einstein's equations, ${ R^{(1)}}_{\mu\nu}+ \epsilon  \,{ R^{(2)}}_{\mu\nu}  =0$, leads the weakly nonlinear wave equation for the perturbed metric:
\begin{eqnarray} {\partial^\lambda}_\lambda  h_{\mu\nu} = - 2 \epsilon  \,{ R^{(2)}}_{\mu\nu} + \dots . \label{Ricci}
\end{eqnarray}

The limit $\epsilon\to 0$ provides  up to first order the linear wave equation for the perturbed metric $h_{\mu\nu}$, which  may be written in general by a plane wave  normal mode  expansion \cite{weinberg}
\begin{eqnarray}
h_{\mu \nu } =\sum_{s,\sigma} \int  \frac{1}{ \sqrt{2\omega_k}} A^{s}_{{\bm k},\sigma}(t)  e^{(s)}_{\mu\nu}({\bm k},\sigma)  e^{i {\bm k}\cdot {\bm x}  }   d^3 {\bm k}. \label{h}
\end{eqnarray} 
Here $A^{s}_\sigma$ correspond  to the Fourier amplitudes of the perturbed metric, it has units of $(\text{length}^3/\sqrt{\text{time}}$). Moreover, $ e^{(s)}_{\mu\nu}({\bm k},\sigma) $  is the polarization normalized tensor (${e^{(s)}}^{\mu\nu}({\bm k},\sigma) e^{(-s)}_{\mu\nu}({\bm k},\sigma)=1$). On the other hand, $\bm k$ and $k = \sqrt{|{\bm k}^2|}$ denote the 3-D wave vector and its modulus, and, $\omega_k= c k $ is the dispersion relation ($c$ the speed of light). The index $\sigma$  labels the two polarization states $\times$ and $+$, and the index $s=\pm 1$ de number of degrees of freedom of each wave corresponding to the two traveling wave solutions. Indeed, the first order, ${ R^{(1)}}_{\mu\nu}=0$ provides a second order differential equation (o.d.e) for the amplitudes: $ \ddot  A^{s}_{{\bm k},\sigma}(t)  + \omega_k^2   A^{s}_{{\bm k},\sigma}(t) =0$, leading to the two independent solutions:
$ A^{s}_{{\bm k},\sigma}(t) \sim e^{\pm i \omega_k t} .$  Because the original fields are real, the amplitudes satisfy $A^{+}_{{\bm k},\sigma} =A_{{\bm k},\sigma}$   and $A^{-}_{{\bm k},\sigma} =A^*_{-{\bm k},\sigma}$, where the $*$ stands for complex conjugated variable.

Up to the next order, the nonlinear wave equation (\ref{Ricci}) provides a set of nonlinear o.d.e's~:
\begin{eqnarray} \dot A^{s}_{{\bm k},\sigma}&=& i s \omega _ k A^{s}_{{\bm k},\sigma} \nonumber\\
& &- i s \epsilon \frac{2 c^2 }{(2\pi)^3 } \frac{  {e^{(-s)}}^{\mu\nu}({\bm k},\sigma)  }{\sqrt{2 \omega_k}  }   \int R^{(2)}_{\mu\nu}  e^{-i {\bm k}\cdot {\bm x}  }  d^3 {\bm x} .\nonumber\\
\label{NLEqn}
\end{eqnarray} 
The perturbed metric  (\ref{h}) has to be substituted in $  R^{(2)}_{\mu\nu} $ at the r.h.s. of (\ref{NLEqn}).

This nonlinear set of o.d.e's  (\ref{NLEqn}) possesses two distinct time scales: a fast oscillation $is\omega_k   A^{s}_{\sigma} $ and the weak quadratic nonlinear oscillation frequency. The fast oscillation maybe removed after a change of variables 
$A^{s}_{{\bm k},\sigma}(t)= a^{s}_{{\bm k},\sigma}(t)e^{i s \omega_k t}.$ Here the explicit time dependence on $a^{s}_{{\bm k},\sigma}(t)$ is a slow time variable. It is important to notice that this change of variables makes the original fields (\ref{h}) to be covariant in the fast variable after replacing  $A^{s}_{{\bm k},\sigma}(t)  e^{i {\bm k}\cdot {\bm x}  } $  in (\ref{h}) by  $a^{s}_{{\bm k},\sigma}(t)   e^{i k^{(s)}_\lambda x^\lambda}$, here the quadri-vector  $k^{(s)}$ reads $k^{(s)}= (-s \omega_k, {\bm k})$.  Using this notation, one computes
the r.h.s. term of (\ref{NLEqn}) that includes $  \frac{  {e^{(-s)}}^{\mu\nu} }{\sqrt{2 \omega_k}  }R^{(2)}_{\mu\nu}   $. After a long, but direct, calculation one gets the pertinent result for the nonlinear terms of the r.h.s. of (\ref{NLEqn}).

 As a sake of simplicity, for the following we shall present only one polarization state, say we take $a_+ =0$ and $a_\times\neq 0$. In this simplified case the equations for the amplitude, $a^s_{{\bm k},\times}(t)$, reads:

   \begin{eqnarray}
\dot a^s_{{\bm k},\times}(t) &=&  \epsilon   \sum_{s_1,s_2}\frac{1}{(2\pi)^3}  \int   L^{s s_1 s_2}_{ {\bm k}  {\bm k}_1  {\bm k}_2} \,  a^{s_1}_{{\bm k}_1,\times}(t)  \,a^{s_2}_{{\bm k}_2,\times}(t)  \nonumber \\
& & \times e^{i (s_1 \omega_{ k_1}+s_2 \omega_{ k_2} - s \omega_{ k})t  }  \delta^{(3)} ({\bm k}_1+{\bm k}_2 -{\bm k} )\, d^3 {\bm k}_1d^3 {\bm k}_2 ,\nonumber\\ \label{ZakhEqn}
\end{eqnarray} 
with the wave interaction amplitude
    \begin{eqnarray} L^{s s_1 s_2}_{ {\bm k}  {\bm k}_1  {\bm k}_2} &= &-is \frac{c^2}{ ( 8 \omega_k \omega_1 \omega_2)^{1/2} }   \left( \sum_{r=s,s_1,s_2} {k^{(r)}}_\mu {k^{(r)}}_\nu   \right)  \nonumber\\ & & \times {\mathcal P}_{s12}  \left[  \frac{1}{4}  {e^{(s)}}^{\mu\nu} {e^{(1)}}^{\lambda\rho}  {e^{(2)}}_{\lambda\rho}  -    {e^{(s)}}_{\lambda\rho}   {e^{(1)}}^{\lambda{\nu}}    {e^{(2)}}^{{\mu}\rho} \right] .\nonumber\\
  \label{L123}
 \end{eqnarray}
Here, ${\mathcal P}_{s12} $ stands for the cyclic permutations of $s$, 1 and 2. In (\ref{L123}) we used the short hand notation $\omega_i = \omega_{k_i}$,  ${e^{(i)}}_{\mu\nu}= {e^{(s_i)}}_{\mu\nu}({\bm k}_i,\sigma_i) $ and ${k^{(s_i)}}_\nu=  (-s_i \omega_{k_i}, {\bm k}_i)$. Moreover, the tensor $ L^{s s_1 s_2}_{ {\bm k}  {\bm k}_1  {\bm k}_2} $ maybe arranged in a fully symmetric structure \cite{newell}. 
Notice that the interaction coefficients scales as $ L^{s s_1 s_2}_{ {\bm k}  {\bm k}_1  {\bm k}_2} \sim \sqrt{\omega_k}\sim \sqrt{c k}$.  
We underline, that the detailed complex structure of the interaction coefficients $L^{s s_1 s_2}_{ {\bm k}  {\bm k}_1  {\bm k}_2}$ does not have any consequence in the results of this letter. The explicit computation is not required in many general results of the wave turbulence theory.  Equations (\ref{ZakhEqn}) and (\ref{L123}) are the starting point of our approach.

{\it Wave turbulence theory.-}
The wave turbulence theory provides an infinite hierarchy equations for the second order cumulant of the Fourier amplitudes 
$$ \left<a_{{\bm k}_1,\times} (t) \,  a^{*}_{{\bm k}_2,\times}  (t)\right> = n _{{\bm k}_1}(t)  \delta^{(3)} ( {\bm k}_1-{\bm k}_2) $$
 in terms of third order cumulants, $ \left< a^{s_1}_{{\bm k}_1,\times} a^{s_2}_{{\bm k}_2,\times} a^{s_3}_{{\bm k}_3,\times}\right>$. The dynamics of the third order cumulant depends on the fourth order cumulant, etc \cite{benney,newell,ZakhBook,NazarenkoBook}. It is shown that because of the non trivial resonant condition, in the limit $\epsilon \to 0$ there exists an asymptotic closure that ensures that the third, fourth, and higher order cumulants depend explicitly of the second order cumulant. 
 Hence the long time statistical properties of a random fluctuating wavy system possess a natural asymptotic closure because of the weakly nonlinear wave interaction. 
 
The final kinetic equation for the wave spectrum of the $\times$ polarization reads (here we use preferentially the methodology of Ref. \cite{newell})

\begin{eqnarray}  
\frac{d}{dt} n_{\bm k}   &=& \epsilon^2 {\mathcal C}oll[n] =
  \frac{\epsilon^2}{2(2\pi)^2 } \sum_{s_1\, s_2} \int d^3 {\bm k}_1  d^3 {\bm k}_2\, |L^{+ s_1 s_2}_{ {\bm k}  {\bm k}_1  {\bm k}_2}|^2    \nonumber\\
& \times&  n_{{\bm k}}    n_{{\bm k}_1}  n_{{\bm k}_2} \left[  \frac{1}{ n_{{\bm k}}  } - \frac{s_1}{ n_{{\bm k}_1}  } -  \frac{s_2}{ n_{{\bm k}_2}  }  \right]   \nonumber\\
& \times& \delta( \omega_k -s_1 \omega_{k_1}-s_2\omega_{k_2} )   \delta^{(3)} ( {\bm k}-  {\bm k}_1- {\bm k}_2 )  .
\label{KineticEqn}
\end{eqnarray}
Here the sum $\sum_{s_1\, s_2}$  produces formally four terms, but one of these terms ($s_1=s_2=-1$) is not resonant, thus it does not contribute to the kinetics. Other cases includes explicitly all possible 3-wave processes, that is a decaying process of two waves into one $2\to 1$, that and  the gain process $1\to 2$.

As for the usual Boltzmann equation, the kinetic equation (\ref{KineticEqn}) conserves
 the total 
the  energy density   (per unit volume)
\begin{eqnarray} {\mathcal E} _{GW} = \frac{c^2}{16\pi G} \int  \omega_k  n_k d^3{\bm k},
\label{GravWaveEnergy}
\end{eqnarray} 
and the momentum density ${\bm S}_{GW} =   \frac{c^2}{16\pi G} \int {\bm k} n_{{\bm k}}  \, d^3{\bm k} $.
Here $G$ is the gravitational constant \cite{foot1}.

Kinetic equation (\ref{KineticEqn}) exhibits a 
$H$-theorem: let be ${ \mathcal S}(t)=\int  \log[n_{{\bm k}}] \, d^3 {\bm k}$ the non-equilibrium entropy, then $d \mathcal S/dt \ge 0$, for increasing time.
The kinetic equation (\ref{KineticEqn}) describes thus an irreversible evolution 
of the wave-spectrum towards equilibrium, this is the Rayleigh-Jeans  distribution or spectrum \cite{RayleighJeans}  which 
reads (for the zero momentum density situation)
\begin{equation}
n^{RJ}_{\bm k}= \frac{T}{\omega_k },
\label{RJ}
\end{equation}
where $T$ is called, by analogy with thermodynamics, the 
temperature (with units of $\text{length}^3/\text{time}^2$)  which is naturally related to the initial energy by 
${\cal E}_{GW} =   \frac{c^2}{16\pi G}   \int \omega _k n^{RJ}_{k} d^2\bm k=     \frac{c^2}{16\pi G}  T  \int  d^3\bm k$. The quantity $ \int  d^3\bm k$ is the number of degrees of freedom per unit volume. Therefore each degree of freedom takes the same energy $T c^2 /16 \pi G $. Naturally, for an infinite system this energy diverges. This classical
 Rayleigh-Jeans catastrophe is always suppressed due to some physical cut-off something not included in Einstein's theory.

{\it Kolmogorov Spectrum.-} In addition of the Rayleigh-Jeans distributions, non-equilibrium solutions can also arise. They have a major importance in the non equilibrium process for the energy transfer among scales. Those solutions can be guessed via a dimensional analysis argument but, as it was shown by Zakharov \cite{zakplasma67,ZakhBook}, they are exact solutions of the kinetic equation.

After Eqn. (\ref{GravWaveEnergy}), the spectral density energy per wave number (with units of $\text{mass}/\text{time}^2$) reads 
$E_k =  \frac{c^2}{4G} \omega_k  k^2 n_k $, so that $ {\mathcal E} _{GW}  = \int E_k \, dk$, then one defines the energy flux $P(k)$ such a that 
$\frac{d}{dt} E_k  = - \frac{d}{dk} P(k).
$
The energy flux $P(k)$ is directly computed after an integration of the r.h.s.  collisional term of equation (\ref{KineticEqn}) and it  has units of $\text{mass}/(\text{length}\, \text{time}^3)$. Assuming a constant energy flux by pure dimensional analysis it is expected that the energy cascade becomes:
\begin{equation}
 E_k = K \left(\frac{P c^3}{G}\right)^{1/2} \frac{1}{k^{1/2}}  ,
\label{KZ}
\end{equation}
here $K$ is a number that will be computed exactly.

This spectrum predicts the existence of  a Kolmogorov turbulent energy cascade of gravitational waves emitted randomly by various sources over the space-time. Moreover, as already said, the this power spectrum is an exact solution that vanishes exactly the r.h.s. collisional term of eqn. (\ref{KineticEqn}).

For the special case of an isotropic power law solution of the form $n_k = A k^{-\alpha}$ one gets that the collisional term of (\ref{KineticEqn}) reads
 $
{\cal C}oll[n_k] =A^2 k^ {3-2 \alpha} I(\alpha),
$
where $I(\alpha) $ is a function of the exponent $\alpha$, and depends explicitly on the form of the interaction coefficients $|L^{s s_1s_2}_{ {\bm k} {\bm k}_1  {\bm k}_2}|^2$.

Moreover, following the Zakharov method, reproduced in \cite{ZakhBook,NazarenkoBook}, one may treat separately the collisional integral (\ref{KineticEqn}) the three different  wave interactions processes $1\to 2$ and $2\to 1$. For the process $2\to 1$, one realizes the conformal transformation: $ k_1= k^2/\tilde k_1  \,   k_2= \tilde k_2 k/\tilde k_1 $ after 
 re-arranging the terms one recovers an integrand that is proportional to the $1\to 2$ process. One gets at the end:

 \begin{eqnarray} I(\alpha) & =& \int _0^\infty \int _0^\infty\frac{S_{1,\kappa_1,\kappa_2}}{ \kappa_1^\alpha  \kappa_2^{\alpha}}\left( 1  -  \kappa_1^{\alpha} - \kappa_2^{\alpha}\right) \nonumber \\ & & \times \left( 1  - \kappa_1^{\beta}- \kappa_2^{\beta}\right) \delta( 1-   \kappa_1 - \kappa_2) \, d\kappa_1\, d\kappa_2,
\nonumber\end{eqnarray} 
 where $\beta = 2\alpha -6$ and 
$
 S_{kk_1k_2} = \frac{k_1 k_2  }{8\pi c k }   |L^{ss_1s_2}_{ {\bm k} { \bm k}_1^*  {\bm k}_2^*}|^2, $ results after the integration of $|L^{s s_1s_2}_{  {\bm k} {\bm k}_1  {\bm k}_2}|^2 \delta^{(3)} ( {\bm k}-  {\bm k}_1- {\bm k}_2 )$ over the solid angles of ${\bm k}_1$ and ${\bm k}_2$. The wave vectors ${ \bm k}_1^* $ and $ {\bm k}_2^*$ are evaluated at the angles allowed by the $\delta$-function condition. $S_{kk_1k_2}$ does not depend on the signs $s_i$ and it could be symmetrized in $k, k_1,k_2 $. Finally, because $L \sim \sqrt{c k}$, the degree of homogeneity of  of the final scattering matrix is $S_{kk_1k_2} \sim k^2$.

As expected, $I(\alpha)$ vanishes for $\alpha =1$ (the Rayleigh-Jeans distribution) and for $\beta=1$, that is  $\alpha=7/2$, which corresponds to the KZ stationary solution.  In general, the energy flux depends explicitly on the function  $I(\alpha) $:
$$ P = -\frac{c^3}{4 G} A^2 \frac{I(\alpha) }{7-2\alpha } k ^{7-2\alpha}.
$$
Because we are looking for stationary solution for the spectral distribution, the flux must be wave number independent, which happens just for $\alpha = 7/2$. Moreover the constant $A$ is completely fixed by the flux (here $I'(7/2)= I'(\alpha)|_{\alpha = 7/2}$)
$ A =  \left( \frac {8 G  P}{c^3 I'(7/2)}\right)^{1/2},
$
in agreement with previous eqn. (\ref{KZ}) by setting there $K = 1/\sqrt{2 I'(7/2)}$. Finally, we underline that  $I(1)=I(7/2)=0$ and one can prove that $I''(\alpha)\geq 0$, so that in general, $ I'(7/2)>0$. Notice that all these relations hold for any   $|L^{s s_1s_2}_{ {\bm k} {\bm k}_1  {\bm k}_2}|^2$, the explicit interaction only modifies the specific numerical value of $I'(7/2)$.

{\it Discussion.-}
Wave turbulence theory appears to successfully describe the dynamics of random gravitational waves.
The present paper derives a kinetic equation for the irreversible spectral transfer among waves. The resulting kinetic equation is able to determine precisely the evolution of the spectral wave distribution in time. In particular, a single black hole coalescence event creates a disturbance that propagates preserving the total the gravitational energy released. Nevertheless, the energy is redistributed among modes driven the system to an equilibrium.
Mandated by the H-theorem, the Rayleigh-Jeans distribution (\ref{RJ}) appears to be the equilibrium attractor for the wave spectrum. So that the original metric (\ref{h}) power spectrum scales as: $\left< | \hat h_{\mu\nu} |^2 \right> _{RJ} \sim \frac{1}{\omega_k} n^{RJ}_k \sim k^{-2}\sim f^{-2}$, where $f$ is the frequency of wave (because the system is non-dispersive $k\sim f$). This metric is directly related to the measured strain, indeed taking the Fourier transform of the signals released \cite{online} from Fig. 1 of Ref. \cite{abbott}, one readily notices that the power spectrum behaves as $f^{-2}$.


Besides the Rayleigh-Jeans equipartition the system manifests the existence of an energy cascade. Presumably this energy cascade appears as a consequence of the  coalescence of black-holes that are random sources of gravitational waves energy.  This process builds a cascade that transfer the energy from one scale (mode) to another. 
In the case of an energy cascade the power spectrum scales as: $\left< | \hat h_{\mu\nu} |^2 \right> _{KZ}\sim \frac{1}{\omega_k} n^{KZ}_k \sim k^{-9/2}\sim f^{-9/2}$. This behavior is not observed in  Ref. \cite{abbott}.
However, as a consequence of the first observed black-hole merger, Ref. \cite{abbott.Stochastic} suggests an scenario of an stochastic background.
The wave turbulence theory predicts that  spectral dimensionless energy density $\Omega_{GW}$ behaves as $k^{1/2}$, which differs slightly, at least over the pertinent decade, from $k^{2/3}$ the inferred one in Ref. \cite{abbott.Stochastic}.

The same calculations for both polarizations $+$ and $\times$ can be done similarly, all scaling properties are unchanged. It is possible to derive a coupled set of kinetic equations as (\ref{KineticEqn}) involving the spectral densities of $n^\times_{\bm k} $ and $n^+_{\bm k} $, including, in particular, the interactions among the mixed states of polarization. The specific calculations would be presented elsewhere.

Notice that usually the resonant condition obtained in the kinetic equation (\ref{KineticEqn}) requires a subtle treatment. The condition inside the $ \delta( \omega_k - \omega_{k_1}- \omega_{k_2} ) $ leads to  a finite contribution only if all involved wave vectors are co-linear. A careful treatment can be found in Refs. \cite{aucoing,lvov}. 
The final kinetic equation contains only one wave number integration instead of two, because of its simplicity it allows us to study the kinetics, the relaxation to a Rayleigh-Jeans equilibrium and the set-up of a KZ spectrum. This subject would be considered in  separate publication.

We end the present paper commenting on a broader aspects of the present work. The wave turbulence theory provides an alternative picture to the so called Cauchy problem \cite{weinberg} in vacuum. Wave turbulence theory provides a systematic methodology to characterize the long time evolution of some suitable initial value problem (notice that this must be stochastic and one should beware of the initial constrains \cite{weinberg}). The kinetic equation (\ref{KineticEqn}) mandates in general an irreversible dynamics for the spectral wave amplitudes. On the other hand it is well known that Einstein's equation, under some conditions, may produce finite-time singularities \cite{khalat,penrose}. Though apparently a paradox, there is no contradiction about these two approaches, wave turbulence may coexist with singular events which would be responsible of intermittency corrections, leading to corrections to the final spectra. By analogy with nonlinear optics in a focusing media (which is known to produces finite-time singularities) \cite{Dyachenko-92,newell}, it is expected that in gravitational waves an energy cascade is built because of the resonant wave interaction, however the wave dynamics is accompanied by frequent spontaneous singular events. These finite-time singularities occur very fast and at a very small scale ($k \to\infty$), thus these singularity structures are a source of energy at the large wavenumber scale. Next, how is the energy transferred to the small wave numbers? How does it occur without any formal inverse cascade? How does this phenomenology modify  the Kolmogorov spectrum? Can this fully nonlinear process give information about the gravitation in extreme situations?  The new era of gravitational astronomy could be a window for these questions.

It is a pleasure to thank A. Gomberoff and D.C. Roberts for interesting discussions on the subject. The author acknowledges the financial support of the CONICYT grant BASAL-CMM. I have made use of data obtained from the LIGO Open Science Center (https://losc.ligo.org), a service of LIGO Laboratory and the LIGO Scientific Collaboration. LIGO is funded by the U.S. National Science Foundation.

 \end{document}